\documentclass[aps,prd,onecolumn,superscriptaddress,amsfont,graphicx,nofootinbib,preprintnumbers,groupedaddress]{revtex4-1}%
\usepackage{color,graphicx,epsfig}
\usepackage{ifpdf}
\usepackage{amsmath}
\usepackage{bm}
\usepackage{color}
\usepackage[english]{babel}
\usepackage{graphicx}%
\usepackage{amsfonts}%
\usepackage{amssymb}
\usepackage{braket}
\usepackage{hyperref}

\usepackage{array}
\newcolumntype{L}[1]{>{\raggedright\let\newline\\\arraybackslash\hspace{0pt}}m{#1}}
\newcolumntype{C}[1]{>{\centering\let\newline\\\arraybackslash\hspace{0pt}}m{#1}}
\newcolumntype{R}[1]{>{\raggedleft\let\newline\\\arraybackslash\hspace{0pt}}m{#1}}

\definecolor{nicered}{rgb}{0.7,0.1,0.1}
\definecolor{nicegreen}{rgb}{0.1,0.5,0.1}
\hypersetup{colorlinks,citecolor= nicegreen,linkcolor= nicered}

\newcommand{\beq}{\begin{equation}}
\newcommand{\eeq}{\end{equation}}
\newcommand{\bea}{\begin{eqnarray}}
\newcommand{\eea}{\end{eqnarray}}

\newcommand{\eventually}[1]{}

\begin{document}
\topmargin -1.0cm
\oddsidemargin -0.8cm
\evensidemargin -0.8cm
\begin{flushright}
ICAS 032/18 \\
\end{flushright}

\def\BuenosAires{~ \\ International Center for Advanced Studies (ICAS)\\
CONICET \& ECyT UNSAM\\ Campus Miguelete, 25 de Mayo y Francia, (1650) Buenos Aires, Argentina}

\title{Exercise: Dark Matter as fields that evolve backward in time}

\author{Ezequiel Alvarez} 
\email[Electronic address: ]{sequi@unsam.edu.ar} 
\affiliation{\BuenosAires}

\preprint{}

\begin{abstract}
We do the exercise of thinking the qualitative features of a new model in which the problems of Dark Matter (DM) and the Cosmic Microwave Background (CMB) homogeneity are apparently simultaneously solved.  We consider that DM consists of fields that evolve backward in time and have a positive defined energy to provide the correct observed curvature of space.  We postulate that these fields do not interact with SM fields other than by sharing the same metric, therefore physics in small curvature regimes is causal, without contradictions with available experiments.  In strongly curved space causality is lost and solutions must fulfill given equations.  The homogeneity problem is naturally solved by these DM fields since, evolving from our future to our past, they bring patches that are currently thought to be causally disconnected into thermal equilibrium for $t\gtrsim 0$ by the Big Bang time.  This thermal equilibrium of DM fields in a strongly curved regime must be shared by the metric and our fields, therefore we see these patches with the same CMB temperature. Although we find unlikely that Nature is described by this theory, we find interesting to conclude that there are not obvious logical nor observational inconsistencies and we explore its features.  Hopefully some thoughts and results in this work may be useful for other ideas.

\end{abstract}

\maketitle

~\newline

\section{Introduction}

Dark matter (DM) is the name assigned to hypothetical particles whose effects are seen as curvature of the space due to its presence.  However, it has not been possible yet to measure an interaction of DM with Standard Model (SM) particles.

Since the discovery of DM gravitational effects, there has been proposals of many kinds to describe DM.  Studies of structure formation point out that most DM should be cold \cite{Davis:1985rj,Blumenthal:1984bp}, although also some hot DM \cite{Davis:1992ui,Caldwell:1993kn} could be expected as well.  To explain DM physicists have analyzed a vast spectrum of possibilities, in particular there are strong efforts dedicated to primordial black holes \cite{Carr:1975qj,Green:1997sz,Carr:2016drx}, axions \cite{Nomura:2008ru,Steffen:2008qp}, sterile neutrinos \cite{Boyarsky:2006fg,Hansen:2001zv,Shi:1998km}, WIMPs \cite{Cushman:2013zza}, and ghost condensate \cite{ArkaniHamed:2003uy} among many others.  Moreover, many of these models make the reasonable assumption that not all DM is the same kind of particle and they postulate given scenarios in which, either to be able to observe it, or to be able to perform a calculation, a fraction of the total DM has some specific behaviour.   On the other hand there is a constant and improving experimental side which looks for direct \cite{Aprile:2017iyp,Akerib:2016vxi,Amole:2015pla} and indirect \cite{TheFermi-LAT:2017vmf,Fermi-LAT:2016uux,Cardano:2008zz} signatures of DM interacting with SM particles.  The direct detection experiments is at this point reaching such a sensitivity that in a few years any possible signal would be blocked by the interaction of the solar neutrinos.   Given all these efforts and negative results, at some point one may consider that in fact both sectors are elusive: there is not interaction between DM and SM other than by sharing the same metric.

We consider a model in the direction of DM with forbidden interactions to SM other than through gravity. More precisely we do the exercise of considering a theory in which DM consist of particles evolving backward in time, but with no direct interaction with SM particles.  DM and SM particles only share the metric, therefore their interaction is practically imperceptible unless the space is strongly curved.  Along the next paragraphs we show how this hypothesis is not leading in principle to any contradiction, and it may simultaneously provide a natural solution to the CMB homogeneity problem.  

This work is divided as follows.  In Section 2 we do a qualitative description of the main features of the considered theory.  In Section 3 we consider one possible mathematical framework that contains the main features discussed in Section 2.  In Section 4 we briefly discuss how this theory would affect some known results and which directions we consider relevant to study.  Section 5 contains the conclusions. 

~\newline
\section{Conceptual description of the model and its features}

Motivated by the symmetry of space-time upon non-orthochronus transformations --which change sign of time-- we consider a theory where actually exist fields that go backward in time, with more specific details to be given below.  We refer to these new proposed fields as primed fields and we prime all quantities that have to do with these fields.  Observe that this proposition is in contrast to usual Quantum Field Theory (QFT) where it is shown that exist transformations of the fields that fulfill the time reversal invariance.  

From now on, for the sake of clarity, we will refer to time as a parameter with different values, setting $t=0$ for the Big Bang, and $t=t_0>0$ for present time.  That is, $t$ increases as we grow old. Just as our SM fields are known to be causal in flat space, we suppose that these primed fields are anti-causal in flat-space: what happens at a given time $t_1$ affects what happens at a time $t_2<t_1$ within a light-cone of anti-causality.  To these primed fields to fulfill the DM phenomenology, we must assume that they have a positive energy ($T'^{00}>0$), as it is shown below.

A first feature that one must study in dealing with primed fields that go backward in time is if they generate contradiction to our daily experimentation of causality.  Moreover, one should also analyze if they could potentially generate a paradox, as for instance if out-going products of an experiment could stop the colliding particles at times smaller than at which the reaction occurs.  Or more naif, if one could excite a primed field that kills his/her grandfather when in the crib and thus one is never born.  This kind of contradictions do not occur in this theory because the only way of interacting fields and primed fields is through gravity and, therefore, in our approximately flat space one cannot voluntarily excite a primed field.  One could argue that, in contrast, in a sufficiently curved space a primed field could be excited and therefore loose the causality.  In fact, this will happen, but we have not experimented causality in strongly curved space.  Moreover, in a strongly curved space we could not exist and therefore what we believe that we could do is actually not achievable\footnote{This is similar to the contrast between Classical and Quantum Mechanics.  From our daily experience in Classical Mechanics we believed that we could simultaneously measure position and momentum of any particle, but when going to the quantum regime, we find that this is not true any more.} simply because we cannot live nor send a working machine in such a strongly curved space.  In more informal words, within this theory, causality and free will are a sensation of approximately flat space; an emergent property. 

What we are therefore considering is a theory in which both SM and DM particles live together and are causal and anti-causal, respectively, in approximately flat space.  They only have a mild interaction through gravity in approximately flat space, which causes for instance a different to the expected velocity of stars as a function of distance to the center of the Galaxy, or gravitational lens effects, as is well known for DM.   In strongly curved space this theory predicts that causality is no longer held and that, instead of having that initial conditions produce a given solution for larger times, what we have is that there exists a solution that must fulfill equation of state for all values of the parameter $t$.  That is, fields and primed fields are latched in an extended region of $t$ by the strong curvature and the equations of motions.  A familiar analogy would be the difference of playing dominoes, where each piece depends on the previous one, and solving a puzzle with latching pieces, where exists a solution that depends on all other pieces.

Up to here we have only considered a new theory that just solves the DM problem; with no much gain.  However, this theory would also solve naturally the problem addressed by inflation of why patches that were never causally connected generate CMB with the same temperature.   This is qualitatively explained in the following reasoning.

Since we understand the meaning of causal theories, it is simpler to analyze the DM primed fields in approximately flat space as normal matter that evolves from larger $t$ to smaller $t$.  In this direction of time --which is contrary to our direction-- the primed fields behave causal, and just as our fields in our direction of time.  We therefore expect that, no matter which is their initial state, they increase their region of thermal equilibrium as time decreases.  If we now take two patches in the sky that today ($t_0$) are outside the horizon, but are going to be within the horizon at a larger time, say $t_1>t_0$, then we cannot understand with the SM and without inflation why these patches have the same temperature in the CMB.  However, in this new theory we may understand it if we think that at $t \gtrsim t_1$ primed fields begin to evolve towards $t=0$ (Big Bang) and their region of thermal equilibrium keeps on increasing until that at $t \gtrsim 0$ they end up putting these two patches in thermal equilibrium for the primed fields.  However, at $t\gtrsim 0$ the curvature is very strong and therefore if the DM primed fields are in thermal equilibrium in these two given patches, then also should be the metric, and the solution must also imply thermal equilibrium for our SM fields.  And this would be the qualitative reasoning of why, within this theory, we see today at $t=t_0$ the CMB photons from these two given patches to have the same temperature.

\section{A possible mathematical framework}

We have qualitatively described in the previous paragraphs how a theory with fields that evolve backward in time would simultaneously solve the problems of DM and the homogeneity of the CMB.  We present in the next paragraphs one possible mathematical framework in which such a theory would be realized.

A possible description of fields that evolve backward in time has been proposed by Lee and Wick in Refs.~\cite{Lee:1969fy,Lee:1970iw} based on important results obtained by Wigner in Ref.~\cite{Wigner:1955zz}.  Recently this kind of fields have been proposed \cite{Grinstein:2007mp} as a natural solution to the hierarchy problem in the SM.  In this class of theories it is proposed that the metric has an overall negative sign which yields a Hamiltonian with opposite-sign and, therefore, the time evolution operator becomes $e^{+i H t}$, instead of the usual\footnote{We assume $\eta_{\mu\nu}=\mbox{Diag}(1,-1,-1,-1)$.} $e^{-i H t}$.  However, this class of Lee-Wick theories are not suitable for our model because of two main failures: {\it (i)} They allow interaction of Lee-Wick fields with SM fields; and --more important-- {\it (ii)} Lee-Wick fields generate a negative energy, in particular $T^{00}_{LW} <0$.  This latter would not generate the desired effect of these fields curving the space as regular matter.

We therefore consider the following new mathematical framework to pursue the effects discussed above.

Usual QFT may be understood as the quantization of the harmonic oscillator and the posterior definition of ladder operators which, when acting on the vacuum, generate a Fock space of states.   All this quantization is strictly related to the usual commutation of the operators $x$ and $p$,
\begin{equation}
 H = \frac{1}{2}(x^2 + p^2) \qquad \qquad \qquad [x,p]=i.
\end{equation}
In defining $[x,p]=i$ one is using as the departing point our daily experience.  One may, however, take as the departing point the Hamiltonian and consider that our daily experience, $[x,p]=i$, is just one possibility.  The other possibility comes from the exchanging symmetry $x \leftrightarrow p$ in the Hamiltonian, which can be extended to the quantization step.  In such a case, one may consider that in addition to the usual form of energy there is also present the one that is obtained by the symmetry extension:
\begin{equation}
 H = \frac{1}{2}(x'^2 + p'^2) \qquad \qquad \qquad [p',x']=i.
\end{equation}
Here $x'$ and $p'$ are position and momentum of some other kind of new particle.  Within this new kind of variables, one can define the new operators
\begin{eqnarray}
a' &=& \frac{1}{\sqrt{2}} (x' - i p' ) \\
a'^\dagger &=& \frac{1}{\sqrt{2}} (x' + i p' ) ,
\end{eqnarray}
which have a different sign to the usual definitions.  These new operators satisfy the commutation relations
\begin{equation}
[a', a'^\dagger ] = 1 \qquad \qquad [a',a']=[a'^\dagger, a'^\dagger]=0,
\end{equation}
and therefore are usual ladder operators.  These operators generate a Fock space with states as usual and with energy positive defined by
\begin{equation}
H = \frac{1}{2} ( a' a'^\dagger + a'^\dagger a' ) = a'^\dagger a' + \frac{1}{2} .
\end{equation}

Since the relativistic extension of $[x',p']=-i$ implies that the space representation of the $p'^0$ operator is now $p'^0 = E = -i\partial_t$, in this new framework we have the exactly same results as usual, but with an opposite-sign time evolution operator:
\begin{eqnarray}
|n'\rangle &=& \frac{1}{\sqrt{n!}} (a'^\dagger)^n |0\rangle \\
H &=& a'^\dagger a' + \frac{1}{2}  \\
E_n &=& (n+\frac{1}{2}) = \langle n' | H | n' \rangle > 0 \\
| \Psi'(t) \rangle &=& e^{+ i H t}  | \Psi'(0) \rangle .
\end{eqnarray}
The prime indicates that the particle content of the state obeys the opposite-sign commutation relations. This is understood as if the state $|\Psi' \rangle$ evolves as a normal state if $t$ decreases instead of increases. 

A similar argument as the previous paragraphs, but for canonical quantization of fields yields to quantum primed fields which have an opposite-sign evolution operator, and therefore one can get a Quantum Field Theory consistently developed to be anti-causal: its evolution for $t$ decreasing corresponds to our usual evolution for $t$ increasing.  To actually state that a system will evolve for time decreasing as our systems evolve for time increasing we need to add a boundary condition for the primed fields at large time; that is in our future.  However, there is no need of fine tuning, practically any boundary condition will provide a solution in which as time decreases the primed fields increase the size of their thermal equilibrium horizon.

We do the exercise of considering that these new primed fields exist in the Universe and correspond to the fields that curve the space as DM.   We postulate that an interaction term of primed and SM fields is prohibited because it would not have well defined time evolution and, mainly, because it has not been observed. 

We therefore write the complete action as
\begin{equation}
S = \int {\cal L}_{SM} + {\cal L'}_{DM} + R,
\label{action}
\end{equation}
where all terms are written with the same metric.  Here ${\cal L}_{SM}$ is the normal SM Lagrangian, whereas ${\cal L'}_{DM}$ corresponds to the new primed fields which obey an opposite-sign commutation relationship and therefore its time evolution corresponds to time decreasing.  We do not have arguments to say more about ${\cal L'}_{DM}$, neither about its particle content, nor its group of symmetry.  For the sake of simplicity, and since there are not quantum observed effects insofar, we consider gravity classical along this exercise.

From Eq.~\ref{action} we obtain one of the equations of motion by varying with respect to the metric
\begin{equation}
G^{\mu\nu} = 8\pi G (T^{\mu\nu} + T'^{\mu\nu}) .
\end{equation}
If the space is approximately flat then fields and primed fields only see each other by gravitational astrophysical effects.  Whereas in strongly curved space we expect that equations of motions of fields and primed fields get latched by gravity.

\section{Discussion}

The theory presented up to here does not seem to have contradictions.  The above paragraphs discuss mainly qualitative features.  It would be interesting to perform a more rigorous study as well as to explore further consequences of this theory in different directions.  We point out some directions in the following paragraphs.

{\it Black Holes.}  Given the time evolution of the fields and primed fields, we would expect that as time increases normal fields fall into a black hole and primed fields are expelled from a black hole.  That is, it would be a white hole for primed fields.  Although white holes have been studied \cite{Kallosh:1995yz,Eardley:1974zz}, in this case a black hole would be a simultaneous mixture of both: black and white hole for matter and DM, respectively, which has not been studied yet.  It would be interesting to study its properties and see if there could be some particular behaviour of it that could be tested in the center of our Galaxy.

{\it Information Paradox\,\footnote{Thanks to M.~Szewc for pointing out this.}.}  This theory provides a good framework to attack the black hole information paradox, a qualitative argument runs as follows.  As time increases, normal matter would enter into a black hole, inside gravity is very strong and there will be interaction between normal matter and DM and, as time increases, DM will come out from the black hole. However, this DM will have encoded the information of the in-going normal matter, since there was an interaction in the black hole. 

{\it Gravitational Waves.} If the DM content of the Universe has enough structure to produce violent enough disruptions or merging events such that produce detectable gravitational waves, then these gravitational waves would be seen as spherical in-going waves as time increases.  This is because the event would have occurred with a dynamic of DM where evolution is from larger to smaller times.  Therefore, current gravitational wave detectors as Ligo and Virgo would detect a gravitational wave passing through but with no visible counter-part at all.  In fact, these gravitational waves would correspond to an event in our future.  Depending of whether such an event of gravitational waves with no visible counter-part are possible or not within normal matter, this could be an interesting signature to explore.

{\it Friedmann Robertson Walker.}  Since this solution requires homogeneity and isotropy and DM fields do not violate this hypothesis, then it would still hold the usual history of the Universe.  It would be interesting, however, to investigate what the radiation and matter dominated era would imply in this new framework.  Within this theory the DM relic-density hypothesis should be revised.

{\it Mathematical framework.} It is arguable that the framework presented in previous Section is too simplistic and yields to similar equations as normal fields.  However, the intrinsic difference in time evolution makes the point of what we try to model with this framework.  It would be interesting to study new different mathematical models that could also predict the desired behaviour of DM evolving from larger to smaller times.

There are certainly many other points which would be interesting to discuss within the framework here considered.


\section{Conclusions}

We have performed the exercise of studying features of a theory in which DM consists of particles that evolve from our future to our past.  We have issued a qualitative argument through which this DM would naturally solve the homogeneity of the CMB from patches that are apparently not causally connected.

We have shown that if DM and usual particles only interact by sharing the same metric, then there would not be a break of causality in approximately flat space, as the one where we live and where we have tested causality insofar.  In strongly curved spaces causality would be lost and it would be interesting to analyze different features within this new scenario.

Along the work we considered one possible mathematical framework to contain this theory: a QFT where, inspired by the symmetry in exchanging $x \leftrightarrow p$ in the harmonic oscillator Hamiltonian, instead of having only $[x,p]=i$ one also has $[p',x']=i$.   This yields to positive defined energy states and an evolution operator with $t \to -t$.

We have presented a very brief discussion in other related subjects such as Black Holes, the information paradox, Gravitational Waves, and the Friedmann Robertson Walker solution.

Along the above paragraphs we have done the exercise of describing qualitatively the behaviour of a new theory which apparently would simultaneously solve the DM and the homogeneity of the CMB problems.  There are no much more motivations than these two problems to consider this theory, therefore is not likely that Nature would be represented as in the presented picture.  However, we have found interesting to perform this study because a theory of a kind has not been thought in these terms and, surprisingly, there are apparently no contradictions, nor logical, nor from the observational point of view.  In addition, the new theory is in some aspect more symmetric than usual QFT.  Hopefully the different ideas presented in this exercise could be useful for future works, for an alternative mathematical framework, or as departing points for new thoughts.

\section*{Acknowledgments}

Thanks to D.L.~L\'opez Nacir, L.~Da Rold, I.~Fabre, M.~Szewc and participants of the workshop {\it Voyages Beyond the SM II} for very useful discussions on the subject of this work.  \eventually{Thanks A.~Yupanqui for suggesting the main idea in this work.}  This work was supported by CONICET and ANPCyT PICT 2013-2266.

\bibliography{biblio}

\end{document}